\input ./style/arxiv-general.cfg
\documentclass[aoas,MSNbibl,nameyear,seceqn,dvips]{arximspdf}
\makeatletter
   \@ifpackageloaded{graphicx}{}{\usepackage{graphicx}}
\makeatother
\usepackage{algorithm}
\usepackage{mathrsfs}

\doi{10.1214/15-AOAS844}
\volume{9}
\issue{3}
\pubyear{2015}
\firstpage{1571}
\lastpage{1600}
\docsubty{FLA}

\makeatletter
\def\widehat{\hat}

\newtheorem{theorem}{Theorem}[section]
\newtheorem{lemma}{Lemma}[section]
\makeatother

\begin{document}
\begin{frontmatter}

\title{Network assisted analysis to reveal the genetic basis of
autism\thanksref{T1}}
\runtitle{Network assisted analysis of autism risk}

\thankstext{T1}{Supported in part by NIH Grants U01MH100233 and R37 MH057881.}

\begin{aug}
\author[A]{\fnms{Li} \snm{Liu}\corref{}\ead[label=e1]{liulipku@gmail.com}},
\author[A]{\fnms{Jing} \snm{Lei}}
\and
\author[A]{\fnms{Kathryn} \snm{Roeder}}
\runauthor{L. Liu, J. Lei and K. Roeder}
\affiliation{Carnegie Mellon University}
\address[A]{Department of Statistics\\
Carnegie Mellon University\\
5000 Forbes Avenue\\
Baker Hall 13\\
Pittsburgh, Pennsylvania 15213\\
USA\\
\printead{e1}}
\end{aug}

%
\received{\smonth{3} \syear{2015}}

%
\begin{abstract}
While studies show that autism is highly heritable, the nature of the
genetic basis of this disorder remains illusive. Based on the idea that
highly correlated genes are functionally interrelated and more likely
to affect risk, we develop a novel statistical tool to find more
potentially autism risk genes by combining the genetic association
scores with gene co-expression in specific brain regions and periods of
development. The gene dependence network is estimated using a novel
partial neighborhood selection (PNS) algorithm, where node specific
properties are incorporated into network estimation for improved
statistical and computational efficiency. Then we adopt a hidden Markov
random field (HMRF) model to combine the estimated network and the
genetic association scores in a systematic manner. The proposed
modeling framework can be naturally extended to incorporate additional
structural information concerning the dependence between genes. Using
currently available genetic association data from whole exome
sequencing studies and brain gene expression levels, the proposed
algorithm successfully identified 333 genes that plausibly affect
autism risk.
\end{abstract}

%
\end{frontmatter}

\section{Introduction}\label{sec1}

Autism spectrum disorder (ASD), a neurodevelopmental disorder, is
characterized by impaired social interaction and restricted, repetitive
behavior. Genetic variation is known to play a large role in risk for
ASD [\citet{Gaugler:2014,Klei:2012}], and yet efforts to identify
inherited genetic variation contributing to risk have been remarkably
unsuccessful [\citet{Anney:2012,Liu:2013}]. One explanation for this
lack of success is the large number of genes that appear to confer risk
for ASD [\citet{Buxbaum:2012}]. Recent studies estimate this number to
be near 1000 [\citet{Sanders:2012,He:2013}].

The advent of next generation sequencing and affordable whole exome
sequencing (WES) has led to significant breakthroughs in ASD risk gene
discovery. Most notable is the ability to detect rare de novo mutations
(i.e., new mutations) in affected individuals. These studies examine
ASD trios, defined as an affected child with unaffected parents, to
determine rare mutations present in the affected child, but not in the
parents. A fraction of these mutations cause loss of function (LoF) in
the gene. And when these rare damaging events are observed in a
particular gene for multiple ASD trios, it lends strong evidence of
causality [\citet{Sanders:2012}]. While this approach has revolutionized
the field, the accumulation of results is slow, relative to the size of
the task: to date, analysis of more than two thousand ASD trios has
identified less than two dozen genes clearly involved in ASD risk [\citet
{Iossifov:2012,Kong:2012,Neale:2012}, \citeauthor{ORoak:2011}
(\citeyear{ORoak:2011,ORoak:2012a}), \citet{Sanders:2012,Willsey:2013,DeRubeis:2014}]. Extrapolating from these
data suggest that tens of thousands of families would be required to
identify even half of the risk genes [\citet{Buxbaum:2012}]. At the same
time, a single de novo LoF (dnLoF) event has been recorded for more
than 200 genes in the available data. Probability arguments suggest
that a sizable fraction of these single-hit genes are ASD genes [\citet
{Sanders:2012}], indicating that genetic data are already providing
partial information about more ASD genes. Thus, there is an urgent need
to advance ASD gene discovery through the integration of additional
biological data and more powerful statistical tests.

The large number of genes with rare coding mutations identified by
exome sequencing presents an opportunity for the next wave of
discoveries. Fortunately a key element in the path forward has recently
been identified. ASD-related mutations have been shown to cluster
meaningfully in a gene network derived from gene expression in the
developing brain---specifically during the mid-fetal period in the
frontal cortex [\citet{Willsey:2013}]. These results support the
hypothesis that genes underlying ASD risk can be organized into a much
smaller set of underlying subnetworks [\citet
{Ben-David:2012,Willsey:2013,Parikshak:2013}]. This leads to the
conjecture that networks derived from gene expression can be utilized
to discover risk genes. \citet{Liu:2014} developed DAWN$_\alpha$, a
statistical algorithm for ``Detecting Association With Networks'' that
uses a hidden Markov random field (HMRF) model to discover clusters of
risk genes in the gene network. Here we present DAWN, a greatly
improved approach that provides a flexible and powerful statistical
method for network assisted risk gene discovery.

There are two main challenges to discovery of ASD genes: (1) weak
genetic signals for association are spread out over a large set of
genes; and (2) these signals are clustered in gene networks, but the
networks are very high dimensional. Available data for network
estimation are extremely limited, hence the dimension of the problem is
orders of magnitude greater than the sample size. A weakness of
DAWN$_\alpha$ lies in the approach to gene network construction. The
algorithm is based on discovering gene modules and estimating the edges
connecting genes within a module based on the pairwise correlations. In
contrast, DAWN estimates the conditional independence network of the
genes under investigation. It achieves this goal utilizing a novel
network estimation method that achieves a dimension reduction that is
tightly linked to the genetic data. Our approach to network assisted
estimation is based on three key conjectures: (i) autism risk nodes are
more likely to be connected than nonrisk nodes;
(ii) by focusing our network reconstruction efforts on portions of the
graph that include risk nodes we can improve the chance that the key
edges in the network that connect risk nodes are successfully
identified and that fewer false edges are included; and
(iii) the HMRF model will have greater power to detect true risk nodes
when the network estimation procedure focuses on successfully
reconstructing partial neighborhoods in the vicinity of risk nodes.

The remainder of this paper is organized as follows. Section~\ref{sec:Background} presents data and background information. Section~\ref{sec:methods} presents the main idea of our testing procedure within a
graphical model framework. First, we develop an algorithm for
estimating the gene interaction network that integrates node-specific
information. Second, we describe the HMRF model. Third, we extend our
model to include the directed network information. Last, we develop in
theory to motivate why our network estimation procedure is more precise
when node-specific information is integrated. In Section~\ref{sec:simu}
simulation experiments compare our approach with other network
estimation algorithms. In Section~\ref{sec:data} we apply our procedure
to the latest available autism data.

\section{Background and data}
\label{sec:Background}

\subsection{Genetic signal}\label{sec2.1}

DAWN requires evidence for genetic association for each gene in the
network. While this can be derived from any gene-based test for
association, a natural choice is TADA, the Transmission And De novo
Association test [\citet{He:2013}]. For this investigation, TADA scores
were calculated using WES data from seventeen distinct sample sets
consisting of 16,098 DNA samples and 3871 ASD cases [\citet
{DeRubeis:2014}]. Using a gene-based likelihood model, TADA produces a
test statistic for each gene in the genome. Based on these data, 18
genes incurred at least two dnLoF mutations and 256 incurred exactly
once. Any gene with more than one dnLoF mutation is considered a ``high
confidence'' ASD gene and those with exactly one are classified as a
``probable'' ASD genes due to the near certainty ($>$99\%) and
relatively high probability ($>$30\%) the gene is a risk gene,
respectively [\citet{Willsey:2013}]. Based on TADA analysis of all genes
covered by WES, 33 genes have false discovery rate (FDR) $q$-values
$<$10\% and 107 have $q$-values $<$30\%. Thus, in total, this is a rich
source of genetic data from which to make additional discoveries of ASD
risk genes and subnetworks of risk genes.



\subsection{Gene networks}\label{sec2.2}

The major source of data from which to infer the gene--gene interaction
network is gene expression levels in specific tissues, which are
obtained by high throughput microarray techniques. Using the BrainSpan
transcriptome data set [\citet{Kang:2011}], Willsey et al. (\citeyear{Willsey:2013})
examined the coexpression patterns across space and time of genes with
at least one dnLoF mutation. The data originate from 16 regions of the
human brain sampled in 57 postmortem brains ranging from 6 weeks
postconception to 82 years of age. By identifying the region and
developmental period of the brain in which ASD genes tend to cluster,
their investigation confirmed that gene expression networks are
meaningful for organization and inter-relationships of ASD genes.
Specifically, they identified prefrontal and motor-somatosensory
neocortex (FC) during the mid-fetal period as the most relevant
spatial/temporal choice. While each brain is measured at only one point
in time, combining gene expression from the five frontal cortex regions
with the primary somatosensory cortex, multiple observations can be
obtained per sample. Nevertheless, the sample size was very small: for
example, for fetal development spanning 10--19 weeks post-conception, 14
brains, constituting 140 total samples, were available from which to
determine the gene network.

Another type of network is the gene regulation network, which is a
directed network. By studying the ChIP-chip data or the ChIP-seq data,
one can obtain which genes are regulated by particular transcription
factors (TFs). Many available gene regulation networks have already
been studied and integrated into a large database called ChEA [\citet
{Lachmann:2010}]. But this kind of network is far from complete. Here
we incorporate the TF network for a single gene (\textit{FMRP}) to
illustrate how this type of information might be utilized in the hunt
for ASD risk genes.

\subsection{Network estimation}\label{sec2.3}

To estimate the gene co-expression network by expression levels, in
general, there are three types of approaches. The most straightforward
way is to apply a correlation threshold: the connectivity of two genes
is determined by whether the absolute correlation is larger than a
fixed threshold. This is the approach taken in the popular systems
biology software tool known as Weighted Gene Co-expression Network
Analysis (WGCNA) [\citet{Langfelder:2008}]. This tool is frequently used
to discover networks of genes, or modules, with high coexpression. The
DAWN$_\alpha$ algorithm used this principle to construct a gene
correlation network [\citet{Liu:2014}]. Using WGCNA, modules were formed
based on the dendrogram with the goal of partitioning genes into highly
connected subunits. Next, to generate a relatively sparse network
within each module, genes with very high correlation were clustered
together into multi-gene supernodes. The motivation for pre-clustering
highly correlated genes as supernodes was to create a network that is
not dominated by local subsets of highly connected genes. By grouping
these subsets of genes into supernodes, the broader pattern of network
connections was more apparent. Finally, the gene network was
constructed by connecting supernodes using a correlation threshold.

A major innovation of the DAWN algorithm developed in this paper is a
more efficient network estimation method with better
statistical interpretation.
Constructing a network based on correlations has two advantages: it is
computationally efficient and the edges can be estimated reliably using
a small sample.
In contrast, the conditional independence network is sparser and has
greater interpretability, but it is much harder to estimate. Assuming
that the gene expression levels follow a multivariate normal
distribution, the conditional independence can be recovered by
estimating the support of the inverse covariance matrix of the
expression data. One approach is to estimate the inverse covariance
matrix directly using penalized maximum likelihood approaches [\citet
{friedman:08,cai:2011,cai:2012,ma:13}]. Alternatively, the neighborhood
selection method is based on sparse regression techniques to select the
pairs of genes with nonzero partial correlations. For instance, \citet
{meinshausen06} applied LASSO for the neighborhood selection of each
gene and then construct the adjacency matrix by aggregating the nonzero
partial correlation obtained for each regression. \citet{peng:09}
proposed a joint sparse regression method for estimating the inverse
covariance matrix. A challenge for both the neighbor selection method
and the maximum likelihood approach is that the number of expression
samples available is two orders of magnitude smaller than the number of
genes. In most applications that utilize LASSO-based methods this
challenge is diminished by simply estimating the gene network for
several hundred genes. For example, \citet{Tan:2014} use a sample of
size 400 to estimate a gene network for 500 genes. For this application
we wish to explore the full range of genes that might be involved in
risk for autism, and thus we cannot reduce the dimension in a naive manner.
One may also consider inverting an estimated covariance matrix [\citet
{Schafer:2005,Opgen:2007}]. But in high dimensions
the matrix inversion may be too noisy.
DAWN takes a novel approach to dimension reduction to optimize the
chance of retaining genes of interest.

\subsection{Networks and feature selection}\label{sec2.4}
Many previous papers have discussed how to incorporate the estimated
network into the feature selection problems, namely, DAPPLE [\citet
{Rossin:2011}], GRAIL [\citet{Raychaudhuri:2009}] metaRanker [\citet
{Pers:2013}], Hotnet [\citet{Vandin:2011}], VEGAS [\citet{Liu:2010}] and
penalized methods [\citet{Mairal:2013}]. However, none of these methods
control the rate of false discovery.

Motivated by the work of \citet{Li:2010} and \citet{Wei:2008},
DAWN$_\alpha$ applied a HMRF model to integrate the gene network into a
powerful risk gene detection procedure. In principle, this approach
captures the stochastic dependence structure of both TADA genetic
scores and the gene--gene interactions, while being able to provide
posterior probability of risk association for each gene and thus
control the rate of false discovery. In practice, DAWN$_\alpha$ has a
weakness due to the multi-gene nodes that define the networks. This
complication led to several statistical challenges in the
implementation of the algorithm. Notably, a~post-hoc analysis is
required to determine which gene(s) within a multi-gene node are
associated with the phenotype. With DAWN we can capture the strengths
of the natural pairing of the gene network and the HMRF model without
these added challenges.

\section{Methods}
\label{sec:methods}
The TADA scores together with the gene--gene interaction network provide
a rich source of information from which to discover ASD genes. To
obtain useful information from these data sets, we will need to utilize
existing tools and also to develop novel statistical procedures that
can overcome several challenges. Our model incorporates 3 main
features: (1) Based on DAWN$_\alpha$, a HMRF model combines the network
structure and individual TADA scores in a systematic manner that
facilitates statistical inference. (2) To obtain the most power from
this model, we require a sparse estimate of the gene--gene interaction
network, but the sample size is insufficient to yield a reliable
estimate of the full gene network (approximately 100 observations and
20,000 genes). However, based on the form of the HMRF model, it is
apparent that it is sufficient to estimate the sub-network of the
gene--gene interaction network that is particularly relevant to autism
risk. We provide a novel approach to achieving this goal. (3) Finally,
the statistical model efficiently incorporates additional covariates,
for instance, the targets of key transcription factors that may
regulate the gene network, to predict autism risk genes.

Feature two is the most challenging. Under the high-dimensional
setting, the existing network estimation approaches are neither
efficient nor accurate enough to successfully estimate the network. To
optimize information available in a small sample size, we need to
target our efforts to capture the dependent structure between
disease-associated genes and their nearest neighbors. DAWN uses a novel
\textit{partial neighborhood selection} (PNS) approach to attain this
goal. By incorporating node-specific information, this approach focuses
on estimating edges between likely risk genes so that it reduces the
complexity of the large-scale network estimation problem and provides a
disease-specific network for the HRMF procedure.

To incorporate the estimated network into the risk gene detection
procedure, feature one involves simplifying the HMRF model already
developed for DAWN$_\alpha$ to integrate the estimated network and the
genetic data. By applying the proposed model, the posterior probability
of each gene being a risk gene can be obtained based on both the
genetic evidence and neighborhood information from the estimated gene network.

If additional gene dependence information such as targets of
transcription factor networks are available, they can be incorporated
naturally into the risk gene detection procedure so that better power
can be achieved. To this end, for feature three, we extend the Ising
model by adding another parameter to characterize the effect of such
additional dependence information. This allows simple estimation and
inference using essentially the same procedure.


\subsection{Partial neighborhood selection for network estimation}
\label{sec:algorithm}
To estimate a high-dimensional disease-specific gene network with small
sample size data, we propose the partial neighborhood selection (PNS)
method. Let $X_1,\ldots, X_n$ be the samples from $d$-dimensional
Gaussian random variables with covariance matrix~$\Sigma$. Our goal at
this stage is to estimate the support of the inverse matrix of $\Sigma
$, which is an adjacency matrix $\Omega$. To maximize the power of the
follow-up HMRF algorithm, the estimated adjacency matrix should be as
precise as possible. But, given the high dimensionality and the small
sample size, estimating the support of the entire precision matrix is a
very ambitious goal. To overcome this challenge, it has been noted that
ignoring some components of the high-dimensional parameter will lead to
better estimation accuracy [\citet{Bickel:2004}]. Here we follow this
rationale by estimating entries $\Omega(i,j)$ for a set of selected
entries $(i,j)$, and setting $\hat\Omega(i,j)=0$ for other entries.
Such a selective estimation approach will inevitably cause some bias,
as many components of the parameter of interest are assigned a null
value. However, this approach has the potential to greatly reduce the
estimation variance for the selected components, as the reduced
estimation problem has much lower dimensionality. Such a procedure is
particularly useful in situations where some low-dimensional components
of the parameter are more important for subsequent inference. We will
need to choose the zero entries carefully so that the bias is
controlled. Because our ultimate goal is to detect the risk genes
associated with a particular disease, the dependence structure between
risk genes is more essential in the procedure rather than the
dependence between nonrisk genes. Specifically, we target $\Omega(i,j)$
for genes $i$ and $j$
with higher TADA scores and their high correlation neighbors. Such a
choice can be supported by the HMRF model described in Section~\ref{sec:hmrf} as
well as the theoretical results in Section~\ref{sec:theory}.

\begin{algorithm}
\caption{PNS algorithm}
\label{alg:pns}
\begin{enumerate}
\item $p$-value screening: Exclude any nodes with $p$-value $p_i>t$. The
remaining nodes define $S'=\{i: p_i\leq t\}$.
\item Correlation screening: Construct a graph $G'=\{S',\Omega'\}$,
where $\Omega'$ is an adjacency matrix with $\Omega'_{ij}=I\{|\rho
_{ij}|>\tau\}$, where $\rho_{ij}$ is the pairwise correlation between
the $i$th and $j$th node. Then exclude all isolated nodes to obtain
$S=S' \setminus\{j: \sum_{i \in S'}\Omega'_{ij}=0 \}$.
\item Retrieving neighbors: Retrieve all possible first order neighbors
of nodes in $S$ and obtain node set $V$, where $V=S\cup\{j: |\rho
_{ij}|>\tau, i\in S\}$.
\item Constructing graph: Apply \citeauthor{meinshausen06}'s (\citeyear{meinshausen06}) regression-based
approach to select the edges among nodes in $S$ and between nodes in
$S$ and $V/S$ by minimizing the following $d_1$ individual loss
functions separately:
%
\begin{equation}
L_{i}=\frac{1}{2}\biggl\Vert X_i-\sum
_{j\in V,j\neq i}\beta_{ij}X_j\biggr\Vert ^2+
\lambda \sum_{j\in V,j\neq i}|\beta_{ij}|,\qquad i=1,
\ldots,d_1,
\end{equation}
where $d_1$ is the number of nodes in $S$, $\lambda$ is the
regularization parameter.
Then the graph $\Omega$ of $V$ is constructed as $\Omega
_{ij}=1-(1-E_{ij})(1-E_{ji})$, where matrix $E$ is
\[
E_{ij}=\cases{ %
I\bigl\{|
\beta_{ij}|>0\bigr\}, &\quad $i\in S,$
\vspace*{2pt}\cr
0, &\quad $i\notin S.$}
\]
\item Return $G=(V,\Omega)$.
\end{enumerate}
\end{algorithm}


In the PNS algorithm (Algorithm \ref{alg:pns}), the $p$-values for each
gene are utilized as the node-specific information for the network
estimation. In step 1, we start with the \textit{key genes}, $S'$, defined
as those genes with relatively small TADA $p$-values.
In step 2, we further screen the key genes by excluding any elements
that are not substantially co-expressed with any other measured genes.
This step is taken because the upcoming HMRF model is applied to
networks. Genes that are not highly co-expressed with any other genes
are not truly functioning in the network. The resulting set, $S$,
establishes the core of the network.
In the third step we expand the gene set to $V$ by retrieving all
likely neighbors of genes in the set $S$. The likely partial
correlation neighbors of gene $j\in S$ are identified based on the
absolute correlation $|\rho_{ij}|>\tau$. The superset $V$ includes all
likely risk genes and their neighbors, but excludes all portions of the
gene network that are free of genetic signals for risk based on the
TADA scores.
Similar correlation thresholding ideas have been considered in \citet
{Butte:1999,Yip:2006,Luo:2007}.
Then we apply the neighborhood selection method [\citet{meinshausen06}]
for each gene in the set $S$ to decide which genes are the true
neighbors of risk genes. Note that the estimated graph does not contain
possible edges between nodes in $V \setminus S$, but the edges that
link nodes in $V \setminus S$ will not affect the results of our
follow-up algorithm, so it is much more efficient to not estimate those
edges when we estimate the disease-specific network.
In the fourth step we apply the neighborhood selection algorithm to the
subnetwork $V$.

\textit{Setting threshold values in gene screening}.
The PNS algorithm uses two tuning parameters, $t$ and $\tau$, in the screening
stage. The choice of $t$ and $\tau$ shall lead to a good
reduction in the number of genes entering the network reconstruction
step, while keeping most of the important genes. A practical way of
choosing $t$ would be to match
some prior subject knowledge about the proportion of risk genes. In
general, $t$
shall not be too small in order to avoid substantial loss of important genes.
The choice of $\tau$ is more flexible, depending on the size of the
problem and available computational resource. In our autism data the
number of genes is very large, therefore, a relatively large
value of $\tau$ is necessary. The choice of $\tau=0.7$ has been used
for gene correlation thresholding in the literature [see, e.g.,
\citet{Yip:2006,Luo:2007,Willsey:2013}]. In our simulation study, we
find that the performance of PNS is stable as long as $t$ is not overly
small, and is insensitive to
the choice of $\tau$. More details are given in Section~\ref{sec:simulation_evaluation}.

\textit{Choosing the tuning parameter in sparse regression}.
Finding the right amount of regularization in sparse support recovery
remains an open and challenging problem. \citet{meinshausen:10} and \citet
{Liu:2010b} proposed a stability approach to select the tuning
parameter; however, due to the high-dimension-low-sample-size scenario,
the subsampling used in this approach reduces the number of samples to
an undesirable level. \citet{peng:13} proposed selecting the tuning
parameter by controlling the FDR, but the FDR cannot be easily
estimated in this context. \citet{Lederer:2014a} suggested an
alternative tuning-free variable selection procedure for
high-dimensional problems known as TREX.
Graphical TREX (GTREX) extends this approach to graphical models [\citet
{Lederer:2014b}]. Although this approach produced promising results in
simulated data, it relies on subsampling. Consequently, for some data
sets the sample size will be a limiting factor.

A parametric alternative relies on an assumption that the network
follows a power law, that is, the probability a node connects to $k$
other nodes is equal to $p(k)\sim k^{-\gamma}$. This assumption is
often made for gene expression networks [\citet{Zhang:2005}]. To measure
how well a network conforms to this law, assess the square of
correlation $R^2$ between $\log p(k)$ and $\log(k)$:
%
\begin{equation}
R^2=\bigl(\operatorname{corr}\bigl(\log p(k),\log(k)\bigr)
\bigr)^2. \label{rsquare}
\end{equation}
$R^2=1$ indicates that the estimated network follows the power law
perfectly, hence, the larger the $R^2$, the closer the estimated
network is to achieve the scale-free criteria. In practice, the tuning
parameter, $\lambda$, can be chosen by visualizing the scatter plot of
$R^2$ as a function of $\lambda$. There is no guarantee that the power
law is applicable to a given network [\citet{Khanin:2006}], and this
approach will not perform well if the assumption is violated. As
applied in the PNS algorithm, the assumption is that the select set of
genes in $V$ follow the power law. The PNS subnetwork is not randomly
sampled from the full network, as it integrates the $p$-value and the
expression data to select portions of the network rather than random
nodes. It has been noted in the literature [\citet{Stumpf:2005}] that
the scale-free property of degree distribution of a random subnetwork
may deviate from that of the original full network; however, the
deviation is usually small.
We find the scale-free criterion suitable for the autism data sets
considered in this paper. However, the general performance of PNS and
DAWN does not crucially depend
on this assumption, as we demonstrate in the simulation study in
Section~\ref{sec:simulation_evaluation}.

\subsection{Hidden Markov random field model}\label{sec:hmrf}
Gene-based tests such as TADA reveal very few genes with a $p$-value that
passes the threshold for genome-wide significance. However, after
taking the gene interaction network into consideration, we usually find
that some genes with small $p$-values are clustered. The $p$-values of
those genes are usually not significant individually, but this
clustering of small $p$-values in the network is highly unlikely to
happen by chance. To enhance the power to detect risk genes, we adopt a
HMRF model to find risk genes by
discovering genes that are clustered with other likely risk genes.

First we convert the $p$-values to normal $Z$-scores, $Z = (Z_1; \ldots
;Z_n)$, to obtain a measure of the evidence of disease association for
each gene. These $Z$-scores are assumed to have a Gaussian mixture
distribution, where the mixture membership of $Z_i$ is determined by
the hidden state $I_i$, which indicates whether or not gene $i$ is a
risk gene. We assume that each of the $Z$-scores under the null
hypothesis $(I = 0)$ has a normal distribution with mean 0 and variance
$\sigma_0^2$, while under the alternative $(I = 1)$ the $Z$-scores
approximately follow a shifted normal distribution, with a mean $\mu$
and variance $\sigma_1^2$. Further, we assume that the $Z$-scores are
conditionally independent given the hidden indicators $\mathbf
{I}=(I_1,\ldots,I_n)$. The model can be expressed as
%
\begin{equation}
Z_{i} \sim P(I_{i}=0)N\bigl(0,\sigma_0^2
\bigr)+P(I_{i}=1)N\bigl(\mu,\sigma_1^2\bigr).
\label{model}
\end{equation}

The dependence structure reduces to the dependence of hidden states
$I_i$. To model the dependence structure of $I_i$, we consider a simple
Ising model with probability
mass function
%
\begin{equation}
P(\mathbf{I}=\eta)\propto\exp\bigl(b^t\eta+c\eta^t\Omega
\eta\bigr) \qquad\mbox {for all } \eta\in\{0,1\}^n. \label{ising}
\end{equation}

We apply the iterative algorithm (Algorithm \ref{alg:HMRF}) to estimate
the parameters
and the posterior probability of $P(I_i|\mathbf{Z})$.
\begin{algorithm}[t]
\caption{HMRF parameter estimation}
\label{alg:HMRF}
\begin{enumerate}
\item Initialize the states of node $I_i=1$ if $Z_i>Z_{\mathrm{thres}}$ and 0 otherwise.
\item For $t=1,\ldots,T$
\begin{enumerate}[(a)]
\item[(a)] Update $(\hat b^{(t)},\hat c^{(t)})$ by maximizing the pseudo likelihood
\[
\prod_{i}\frac{\exp\{bI_i+cI_i\Omega_{i\cdot}\mathbf{I}\}}{\exp\{
bI_i+cI_i\Omega_{i\cdot}\mathbf{I}\}+\exp\{b(1-I_i)+c(1-I_i)\Omega
_{i\cdot}\mathbf{I}\}}.
\]

\item[(b)] Apply a single cycle of the iterative conditional mode [ICM, \citet
{Besag:1986}] algorithm to update $\mathbf{I}$. Specifically, we obtain
a new $\hat{I}_j^{(t)}$ based on
\[
P\bigl(I_j|\mathbf{Z};\widehat{\mathbf{I}}_{-i},\hat
b^{(t)},\hat c^{(t)}\bigr) \propto f(z_i| \hat
I_i)P\bigl(I_i|\widehat{\mathbf{I}}_{-i},\hat
b^{(t)},\hat c^{(t)}\bigr).
\]

\item[(c)] Update $(\hat\mu^{(t-1)},\hat\sigma_0^{2 (t-1)},\hat\sigma_1^{2 (t-1)})$
to $(\hat\mu^{(t)},\hat\sigma_0^{2 (t)},\hat\sigma_1^{2 (t)})$:
\begin{eqnarray*}
\hat\mu^{(t)}&=&\frac{\sum_i P(I_i=1|\mathbf{Z},\hat b^{(t)};\hat
c^{(t)}) Z_i}{\sum_i P(I_i=1|\mathbf{Z},\hat b^{(t)};\hat c^{(t)})},
\\
\hat\sigma_0^{2 (t)}&=&\frac{\sum_i P(I_i=0|\mathbf{Z},\hat
b^{(t)};\hat c^{(t)}) Z_i^2}{\sum_i P(I_i=0|\mathbf{Z},\hat
b^{(t)};\hat c^{(t)})},
\\
\hat\sigma_1^{2 (t)}&=&\frac{\sum_i P(I_i=1|\mathbf{Z},\hat
b^{(t)};\hat c^{(t)}) (Z_i-\hat\mu^{(t)})^2}{\sum_i P(I_i=1|\mathbf
{Z},\hat b^{(t)};\hat c^{(t)})}.
\end{eqnarray*}
\end{enumerate}
\item Return $(\hat b,\hat c,\hat\mu,\hat\sigma_0^2,\hat\sigma
_1^2)=(\hat b^{(T)},\hat c^{(T)},\hat\mu^{(T)},\hat\sigma
_0^{2 (T)},\hat\sigma_1^{2 (T)})$.
\end{enumerate}
\end{algorithm}

After the posterior probability of $P(I_i|\mathbf{Z},\mathbf{I}_{-i})$
is obtained, we apply Gibbs sampling to estimate the posterior
probability $q_i=P(I_i=0|\mathbf{Z})$. Finally, let $q_{(i)}$ be the
sorted posterior probability in ascending order; the Bayesian FDR
correction [\citet{muller:2006}] of the $l$th sorted gene can be
calculated as
%
\begin{equation}
\operatorname{FDR}_l=\sum_{i=1}^lq_{(i)}\Big/l.
\label{eq:bFDR}
\end{equation}
Genes with FDR less than $\alpha$ are selected as the risk genes.

In summary, the DAWN algorithm (Algorithm \ref{alg:newdawn}) consists of four steps.

\begin{algorithm}[t]
\caption{DAWN algorithm}
\label{alg:newdawn}
\begin{enumerate}
\item Obtain gene specific $p$-values.
\item Estimate the gene network using the PNS algorithm (Algorithm \ref
{alg:pns}).
\item Incorporate the information from steps 1 and 2 into the HMRF
model and estimate the parameters of the HMRF model (Algorithm \ref{alg:HMRF}).
\item Apply the Bayesian FDR correction to determine the risk genes
[equation (\ref{eq:bFDR})].
\end{enumerate}
\end{algorithm}

The HMRF component of DAWN$_\alpha$ is similar in spirit to what is
described here for DAWN, but the implementation in the former algorithm
is considerably less powerful due to multi-gene nodes. DAWN$_\alpha$
cannot directly infer risk status of genes from the estimated status of
the node.

\subsection{Extending the Ising model}\label{sec3.3}
Our framework is general and flexible enough to incorporate additional
biological information such as the TF network information by naturally
extending the Ising model. Under this extended model, we can
incorporate a directed network such as the TF network along with the
undirected network such as the gene co-expression network. From the
microarray gene expression levels, an undirected network could be
estimated based on the PNS algorithm. With the TF network information,
we could also estimate a directed network that indicates which genes
are regulated by specific TF genes. This additional information can be
naturally modeled in the Ising model framework by allowing the model
parameter to be shifted for particular collection of TF binding sites.
The density function of this more general Ising model is as follows:
%
\begin{equation}
P(\mathbf{I}=\eta)\propto\exp\bigl(b\mathbf{1}'\eta+c
\eta^t\Omega\eta +dH'\eta\bigr), \label{gising}
\end{equation}
where $H=(h_1,\ldots,h_n)$ is the indicator of TF binding sites, and
$d>0$ reflects the enhanced probability of risk for genes regulated by TF.

If $d>0$, this indicates that the TF binding site covariate is a
predictor of risk for diseases. To test whether or not $d$ is
significantly larger than zero, we compare the observed statistic $\hat
d$ with $d$ obtained under the null hypothesis of no association.
To this end, we adopt a smoothed bootstrap procedure which simulates
data with the same clustering of the observed genetic signals, but
without an association with the TF binding site.

To simulate $\mathbf{Z}$ from the null model, we first simulate the
hidden states $\mathbf{I}$ from the distribution (\ref{ising}). We
randomly assign initial values of $\mathbf{I}$ to each node in the
network and the proportion of nodes with $I=1$ is $r$, where $r\in
(0,1)$ is a pre-chosen value, for example, 0.1. Then, we apply a
Metropolis--Hastings algorithm to update $I$ until convergence. The full
bootstrap procedure is described in Algorithm~\ref{alg:sim}.

\begin{algorithm}
\caption{}
\label{alg:sim}
\begin{enumerate}
\item Apply the algorithm to model (\ref{ising}) to obtain estimates of
the model parameters.
\item Using the estimated null model, simulate $\mathbf{I}^*$ by the
Metropolis--Hastings algorithm, then simulate $\mathbf{Z}^*$ using
equation (\ref{model}).
\item Using model (\ref{gising}), estimate the parameters for the
simulated data.
\item Repeat steps 2--3 $B$ times, the $B$ copies of estimated $\hat d$
can be used as
a reference distribution of the estimated parameter under the null model.
\item Output the $p$-value $p=\frac{1}{B}\sum_{i=1}^BI\{\hat{d}_i>d\}$.
\end{enumerate}
\end{algorithm}

For presentation simplicity we describe the idea of incorporating
additional subject knowledge into the Ising model for a single TF. The
procedure can be straightforwardly extended to incorporate multiple TFs.
In this case, the Ising model for the hidden vector $\mathbf I$ becomes
\[
P(\mathbf I = \eta)\propto\exp \Biggl(b\mathbf1' \eta+c
\eta'\Omega\eta +\sum_{k=1}^K
d_k H_k' \eta \Biggr).
\]
The bootstrap testing procedure described in Algorithm \ref{alg:sim}
also carries over in an obvious manner to the multiple TF case.

\subsection{More about partial neighborhood selection}\label{sec:theory}
In this section we discuss theory that explains why PNS can more
precisely estimate edges between risk genes. We find that under the
Ising model, nodes with similar properties are more likely to be
connected with each other in the network. Therefore, by utilizing this
property of the Ising model, we can greatly improve the accuracy of
estimating a disease-specific network.

The following theorem suggests that the larger the $Z$-scores are for the
two nodes, the more likely there is an edge connecting those two nodes.
Therefore, it is reasonable to adapt the lasso regression to retrieve
neighbors of only candidate risk genes, which are the genes that have
small $p$-values. This choice is justified because those genes are more
likely to be connected with other genes.

\begin{theorem}\label{thm:main}
Assume that $(\mathbf Z, \mathbf I)$ are distributed according to the
HMRF in equations (\ref{model}) and (\ref{ising}). Assume that
$\Omega$ has independent entries. Let $\Omega'=\{\Omega
_{k_1,k_2},(k_1,k_2) \neq(i,j)\}$.
$\mathscr{A}$ be the set of all possible $\Omega'$. Define $I_i$ and
$I_j$ as the $i$th and $j$th element of $\mathbf{I}$, $\mathbf
{I'}=(I_1,I_2,\ldots,I_d)/\{I_i,I_j\}$,
and $\mathscr{B}$ the set of all possible $\mathbf{I}'$. Then for any
$\Omega' \in\mathscr{A}$ and any $\mathbf{I}' \in\mathscr{B} $,
$P(\Omega_{ij}=1|\mathbf{Z},\Omega',\mathbf{I}')$ is an increasing
function of $Z_i$ and $Z_j$.
\end{theorem}

Theorem~\ref{thm:main} provides some justification for the $p$-value
thresholding in the PNS algorithm. An important condition here is that
$\mathbf I$ is distributed as an Ising model where the conditional
independence is modeled by the binary matrix $\Omega$.
In practice, if $\Omega$ is estimated from some other data source,
then it is possible that $\Omega$ may not be relevant to reflect the
independence structure of $\mathbf I$.
This is not the case in our application, as the gene co-expression
data is collected
from the BrainSpan data for the frontal cortex sampled during the
mid-fetal developmental period because it has been shown
that this space--time-tissue combination is particularly relevant to
autism [\citet{Willsey:2013}].

\begin{pf*}{Proof of Theorem \ref{thm:main}}
Let $k=1$ represent $(I_i,I_j)=(1,1)$, $k=2$ represent
$(I_i,I_j)=(1,0)$, $k=3$ represent $(I_i,I_j)=(0,1)$, and $k=4$
represent $(I_i,I_j)=(0,0)$. Then
\begin{eqnarray*}
&&P\bigl(\Omega_{ij}=1|\mathbf{Z},\Omega',
\mathbf{I}'\bigr)\\
&&\qquad= \sum_{k=1}^4P
\bigl(\Omega _{i,j}=1,(I_i,I_j)=k|\mathbf{Z},
\Omega',\mathbf{I'}\bigr)
\\
&&\qquad=\sum_{k=1}^4P\bigl(
\Omega_{i,j}=1|(I_i,I_j)=k,\mathbf{Z},
\Omega',\mathbf {I'}\bigr)P\bigl((I_i,I_j)=k|
\mathbf{Z},\Omega',\mathbf{I'}\bigr)
\\
&&\qquad=\sum_{k=1}^4P\bigl(
\Omega_{i,j}=1|(I_i,I_j)=k,
\Omega',\mathbf {I'}\bigr)P\bigl((I_i,I_j)=k|
\mathbf{Z},\Omega',\mathbf{I'}\bigr)
\\
&&\qquad=M_1\bigl(\mathbf{I'},\Omega'
\bigr)P_1+\sum_{k=2}^4
M_2\bigl(\mathbf{I'},\Omega'
\bigr)P_k,
\end{eqnarray*}
where $M_i(\mathbf{I'},\Omega')=P(\Omega_{i,j}=1|(I_i,I_j)=k,\Omega
',\mathbf{I'})$ and $P_k=P((I_i,I_j)=k|\mathbf{Z},  \Omega',\mathbf{I'})$.
Taking a derivative of $P(\Omega_{ij}=1|\mathbf{Z},\Omega',\mathbf
{I}')$ with respect to $Z_i$, we have
\begin{eqnarray*}
\frac{\partial P(\Omega_{ij}=1|\mathbf{Z},\Omega',\mathbf
{I}')}{\partial Z_i} &=& M_1\bigl(\mathbf{I'},
\Omega'\bigr) \times\frac{\partial P_1}{\partial
Z_i}+M_2\bigl(
\mathbf{I'},\Omega'\bigr) \times\biggl(
\frac{\partial P_2}{\partial
Z_i}+\frac{\partial P_3}{\partial Z_i}+\frac{\partial P_3}{\partial
Z_i} \biggr)
\\
&=& M_1\bigl(\mathbf{I'},\Omega'\bigr)
\times\frac{\partial P_1}{\partial
Z_i}-M_2\bigl(\mathbf{I'},
\Omega'\bigr) \times\frac{\partial P_1}{\partial Z_i}
\\
&=& \bigl( M_1\bigl(\mathbf{I'},\Omega'
\bigr) - M_2\bigl(\mathbf{I'},\Omega'\bigr)
\bigr)\times\frac
{\partial P_1}{\partial Z_i}.
\end{eqnarray*}
Based on Lemma~\ref{lemma:1}, we have $M_1(\mathbf{I'},\Omega')-
M_2(\mathbf{I'},\Omega')>0$. Based on Lemma~\ref{lemma:2}, we have
$\frac{\partial P_1}{\partial Z_i}>0$. Thus, we obtain
$\frac{\partial P(\Omega_{ij}=1|\mathbf{Z},\Omega',\mathbf
{I}')}{\partial Z_i}>0$, and $P(\Omega_{ij}=1|\mathbf{Z},\Omega',\mathbf
{I}')$ is an increasing function of $Z_i$. Similarly, we obtain that
$P(\Omega_{ij}=1|\mathbf{Z},\Omega',\mathbf{I}')$ is also an increasing
function of $Z_j$.
\end{pf*}

The theorem above reveals the specific structure of the adjacency
matrix for the network in the Ising model setting. This kind of
adjacency matrix has more edges in the block of risk genes and fewer
edges in the block of nonrisk genes. Thus, given this specific
structure and limited sample size, it is reasonable to focus on
estimating the edges between genes with small $p$-values. Therefore,
under the Ising model, the proposed PNS algorithm is a more precise
network estimation procedure than other existing network estimating
procedures, which all ignore the node-specific information.

\begin{lemma}\label{lemma:1}
Under the same conditions as in Theorem~\ref{thm:main}, for any $\Omega
' \in\mathscr{A}$ and any $\mathbf{I}' \in\mathscr{B} $,
\begin{eqnarray*}
P\bigl(\Omega_{ij}=1|I_i,I_j,
\mathbf{I}',\Omega'\bigr)=\cases{ %
M_1\bigl(\mathbf{I}',
\Omega'\bigr) >P(\Omega_{ij}=1), & \quad $\mbox{if }
I_i=I_j=1,$
\vspace*{2pt}\cr
M_2\bigl(\mathbf{I}',\Omega'\bigr) <P(
\Omega_{ij}=1), &\quad $\mbox{otherwise.}$}
\end{eqnarray*}
\end{lemma}

\begin{pf}
\begin{eqnarray*}
&&\frac{P(\Omega_{ij}=1|I_i=I_j=1,\mathbf{I}',\Omega')}{P(\Omega
_{ij}=0|I_i=I_j=1,\mathbf{I}',\Omega')} \\
&&\qquad= \frac{P(\Omega
_{ij}=1,I_i=I_j=1,\mathbf{I}',\Omega')}{P(\Omega
_{ij}=0,I_i=I_j=1,\mathbf{I}',\Omega')}
\\
&&\qquad=\frac{P(I_i=I_j=1,\mathbf{I}'|\Omega_{ij}=1,\Omega')P(\Omega
_{ij}=1)P(\Omega')}{P(I_i=I_j=1,\mathbf{I}'|\Omega_{ij}=0,\Omega
')P(\Omega_{ij}=0)P(\Omega')}.
\end{eqnarray*}
Let
\[
f\bigl(I_i,I_j,\mathbf{I'}|
\Omega_{ij},\Omega'\bigr)=\exp\bigl(-b^t
\mathbf{I}+c\mathbf {I}^t\Omega\mathbf{I}\bigr),
\]
where $\mathbf{I}_{-i,-j}=\mathbf{I'},\mathbf{I}_i=I_i,\mathbf{I}_j=I_j$.
We define
\begin{eqnarray*}
T_{1}\bigl(\mathbf{I'},\Omega_{ij},
\Omega'\bigr)&=&f\bigl(I_i=1,I_j=1,
\mathbf{I'}|\Omega _{ij},\Omega'\bigr),
\\
T_{2}\bigl(\mathbf{I'},\Omega_{ij},
\Omega'\bigr)&=&f\bigl(I_i=1,I_j=0,\mathbf
{I'}|\Omega_{ij},\Omega'\bigr),
\\
T_{3}\bigl(\mathbf{I'},\Omega_{ij},
\Omega'\bigr)&=&f\bigl(I_i=0,I_j=1,
\mathbf{I'}|\Omega _{ij},\Omega'\bigr),
\\
T_{4}\bigl(\mathbf{I'},\Omega_{ij},
\Omega'\bigr)&=&f\bigl(I_i=0,I_j=0,\mathbf
{I'}|\Omega_{ij},\Omega'\bigr).
\end{eqnarray*}
Then we obtain that
\begin{eqnarray*}
P\bigl(I_i=I_j=1,\mathbf{I}'|
\Omega_{ij}=1,\Omega'\bigr)&=&\frac{T_{1}(\mathbf
{I'},\Omega_{ij}=1,\Omega')}{\sum_{\mathbf{J'}\in\mathscr{B}} \sum_{k=1}^4 T_{k}(\mathbf{J'},\Omega_{ij}=1,\Omega') },
\\
P\bigl(I_i=I_j=1,\mathbf{I}'|
\Omega_{ij}=0,\Omega'\bigr)&=&\frac{T_{1}(\mathbf
{I'},\Omega_{ij}=0,\Omega')}{\sum_{\mathbf{J'}\in\mathscr{B}} \sum_{k=1}^4 T_{k}(\mathbf{J'},\Omega_{ij}=0,\Omega') }.
\end{eqnarray*}
It is easy to show that
\[
T_{1}\bigl(\mathbf{I'},\Omega_{ij}=1
\bigr)=T_{1}\bigl(\mathbf{I'},\Omega_{ij}=0
\bigr)\times \exp\{2c\}
\]
and
\[
T_{k}\bigl(\mathbf{I'},\Omega_{ij}=0
\bigr)=T_{k}\bigl(\mathbf{I'},\Omega_{ij}=1
\bigr),\qquad k=2,3,4.
\]
Therefore,
\begin{eqnarray*}
&&\frac{P(I_i=I_j=1,\mathbf{I}'|\Omega_{ij}=1,\Omega
')}{P(I_i=I_j=1,\mathbf{I}'|\Omega_{ij}=0,\Omega')}
\\
&&\qquad= \frac{ \sum_{\mathbf{J'}\in\mathscr{B}} \sum_{k=1}^4 T_{k}(\mathbf
{J'},\Omega_{ij}=0,\Omega')}{
\sum_{\mathbf{J'}\in\mathscr{B}}  \{\sum_{k=2}^4 T_{k}(\mathbf
{J'},\Omega_{ij}=0,\Omega')+ T_{1}(\mathbf{J'},\Omega_{ij}=0,\Omega
')\times\exp\{2c\}  \} }
\\
&&\quad\qquad{}\times\frac{T_{1}(\mathbf{I'},\Omega_{ij}=0,\Omega')\times\exp\{2c\}
}{ T_{1}(\mathbf{I'},\Omega_{ij}=0,\Omega')}
\\
&&\qquad= \frac{\sum_{\mathbf{J'}\in\mathscr{B}}\exp\{2c\} \sum_{k=1}^4
T_{k}(\mathbf{J'},\Omega_{ij}=0,\Omega')}{\sum_{\mathbf{J'}\in\mathscr
{B}}  \{ \sum_{k=2}^4 T_{k}(\mathbf{J'},\Omega_{ij}=0,\Omega')+
T_{1}(\mathbf{J'},\Omega_{ij}=0,\Omega')\times\exp\{2c\}  \}}
\\
&&\qquad>1 \qquad \mbox{if } c>0.
\end{eqnarray*}
Thus, we obtain
\[
\frac{P(\Omega_{ij}=1|I_i=I_j=1,\mathbf{I}',\Omega')}{P(\Omega
_{ij}=0|I_i=I_j=1,\mathbf{I}',\Omega')} > \frac{P(\Omega
_{ij}=1)}{P(\Omega_{ij}=0)},
\]
which leads to
\[
P\bigl(\Omega_{ij}=1|I_i=I_j=1,
\mathbf{I}',\Omega'\bigr) = M_1\bigl(
\mathbf{I}',\Omega '\bigr) >\frac{P(\Omega_{ij}=1)}{P(\Omega_{ij}=0)+P(\Omega_{ij}=1)}.
\]

Similarly, for any $\Omega' \in\mathscr{A}$ and any $\mathbf{I}' \in
\mathscr{B} $, we obtain
%
\begin{eqnarray}
\label{eq1} &&\frac{P((I_i,I_j)=k,\mathbf{I}'|\Omega_{ij}=1,\Omega
')}{P((I_i,I_j)=k,\mathbf{I}'|\Omega_{ij}=0,\Omega')}\nonumber
\\
&&\quad= \frac{\sum_{\mathbf{J'}\in\mathscr{B}} \sum_{l=1}^4 T_{l}(\mathbf
{J'},\Omega_{ij}=0,\Omega')}{\sum_{\mathbf{J'}\in\mathscr{B}}  \{
\sum_{l=2}^4 T_{l}(\mathbf{J'},\Omega_{ij}=0,\Omega')+ T_{1}(\mathbf
{J'},\Omega_{ij}=0,\Omega')\times\exp\{2c\} \} }
\\
&&\quad<1 \qquad\mbox{when } k=2,3,4 ,
\nonumber
\end{eqnarray}
where $k=2$ means $(I_i,I_j)=(1,0)$, $k=3$ means $(I_i,I_j)=(0,1)$, and
$k=4$ means $(I_i,I_j)=(0,0)$. Then it is easy to obtain
\[
P\bigl(\Omega_{ij}=1|(I_i,I_j)=k,
\mathbf{I}',\Omega'\bigr) = M_k\bigl(
\mathbf {I}',\Omega'\bigr) <\frac{P(\Omega_{ij}=1)}{P(\Omega_{ij}=0)+P(\Omega_{ij}=1)}.
\]
From equation (\ref{eq1}) it is clear that $M_k(\mathbf{I}',\Omega')$
does not depend on $k$, thus
$M_k(\mathbf{I}',\Omega')=M_2(\mathbf{I}',\Omega')$ for $k=2,3,4$.
\end{pf}

Based on Lemma~\ref{lemma:1}, we know that if a pair of nodes has two
risk nodes, then this pair of nodes are more likely to be connected
with an edge than the pairs of nodes with only one risk node or no risk nodes.

\begin{lemma}\label{lemma:2}
Under the same conditions as in Theorem~\ref{thm:main}, for any $\Omega
' \in\mathscr{A}$ and any $\mathbf{I}' \in\mathscr{B} $,
$P(I_i=1,I_j=1 |\mathbf{I}', \Omega',\mathbf{Z})$ is an increasing
function of $Z_i$ and $Z_j$.
\end{lemma}

\begin{pf}
We first derive the conditional probability of $I_i=I_j=1$ given
$\mathbf{I}', \Omega'$ and $\mathbf{Z}$:
\begin{eqnarray*}
P_1&=&P\bigl(I_i=1,I_j=1|
\mathbf{I}',\Omega', \mathbf{Z}\bigr)= P
\bigl(I_i=1,I_j=1,\mathbf{I}',
\Omega',\mathbf{Z}\bigr)/P\bigl(\mathbf{I}',\Omega
',\mathbf{Z}\bigr)
\\
&=&P\bigl(\mathbf{Z}|I_i=1,I_j=1,\mathbf{I}',
\Omega'\bigr)P\bigl(I_i=1,I_j=1,\mathbf
{I}',\Omega'\bigr)/P\bigl(\mathbf{I}',
\Omega',\mathbf{Z}\bigr)
\\
&=&P(Z_i|I_i=1)P(Z_j|I_j=1)P
\bigl(\mathbf{Z}_{-i,-j}|\mathbf {I}'\bigr)P
\bigl(I_i=1,I_j=1,\mathbf{I}',
\Omega'\bigr)\\
&&{}/P\bigl(\mathbf{I}',\Omega',
\mathbf {Z}\bigr)
\\
&=&P(Z_i|I_i=1)P(Z_j|I_j=1)P
\bigl(\mathbf{Z}_{-i,-j}|\mathbf {I}'\bigr)P
\bigl(I_i=1,I_j=1,\mathbf{I}'|
\Omega'\bigr)P\bigl(\Omega'\bigr)\\
&&{}/P\bigl(
\mathbf{I}',\Omega ',\mathbf{Z}\bigr)
\\
&=&P(Z_i|I_i=1)P(Z_j|I_j=1)P
\bigl(\mathbf{Z}_{-i,-j}|\mathbf {I}'\bigr)P
\bigl(I_i=1,I_j=1,\mathbf{I}'|
\Omega'\bigr)\\
&&{}/P\bigl(\mathbf{I}',\mathbf{Z}|
\Omega'\bigr).
\end{eqnarray*}
Similarly, we obtain
\begin{eqnarray*}
P_2&=&P\bigl(I_i=1,I_j=0|
\mathbf{I}',\Omega',\mathbf{Z}\bigr)
\\
&=&P(Z_i|I_i=1)P(Z_j|I_j=0)P
\bigl(\mathbf{Z}_{-i,-j}|\mathbf {I}'\bigr)P
\bigl(I_i=1,I_j=0,\mathbf{I}'|
\Omega'\bigr)\\
&&{}/P\bigl(\mathbf{I}',\mathbf{Z}|\Omega
'\bigr),
\\
P_3&=&P\bigl(I_i=0,I_j=1|
\mathbf{I}',\Omega',\mathbf{Z}\bigr)
\\
&=&P(Z_i|I_i=0)P(Z_j|I_j=1)P
\bigl(\mathbf{Z}_{-i,-j}|\mathbf {I}'\bigr)P
\bigl(I_i=0,I_j=1,\mathbf{I}'|
\Omega'\bigr)\\
&&{}/P\bigl(\mathbf{I}',\mathbf{Z}|\Omega
'\bigr),
\\
P_4&=&P\bigl(I_i=0,I_j=0|
\mathbf{I}',\Omega',\mathbf{Z}\bigr)
\\
&=&P(Z_i|I_i=0)P(Z_j|I_j=0)P
\bigl(\mathbf{Z}_{-i,-j}|\mathbf {I}'\bigr)P
\bigl(I_i=0,I_j=0,\mathbf{I}'|
\Omega'\bigr)\\
&&{}/P\bigl(\mathbf{I}',\mathbf{Z}|
\Omega'\bigr).
\end{eqnarray*}
We further define
\begin{eqnarray*}
C_1&=&P\bigl(\mathbf{Z}_{-i,-j}|\mathbf{I}'
\bigr)P\bigl(I_i=1,I_j=1,\mathbf{I}'|\Omega
'\bigr), \\
 C_2&=&P\bigl(\mathbf{Z}_{-i,-j}|
\mathbf{I}'\bigr)P\bigl(I_i=1,I_j=0,\mathbf
{I}'|\Omega'\bigr),
\\
C_3&=&P\bigl(\mathbf{Z}_{-i,-j}|\mathbf{I}'
\bigr)P\bigl(I_i=0,I_j=1,\mathbf{I}'|\Omega
'\bigr), \\
 C_4&=&P\bigl(\mathbf{Z}_{-i,-j}|
\mathbf{I}'\bigr)P\bigl(I_i=0,I_j=0,\mathbf
{I}'|\Omega'\bigr).
\end{eqnarray*}
Since
\[
P\bigl(I_i,I_j,\mathbf{I}'|
\Omega'\bigr)=P(\mathbf{I}|\Omega)P(\Omega _{ij}=1)+P(
\mathbf{I}|\Omega)P(\Omega_{ij}=0),
\]
it is then clear that $C_k,  k=1,\ldots,4$ is independent with $Z_i, Z_j$. Since
\[
P_1/P_2=\frac{C_1P(Z_j|I_j=1)}{C_2P(Z_j|I_j=0)}=\frac{C_1}{C_2}\exp
\biggl(\mu Z_j-\frac{\mu^2}{2}\biggr),
\]
therefore $P_1/P_2$ is an increasing function of $Z_j$ and independent
with $Z_i$. Similarly, we obtain $P_1/P_3$ is an increasing function of
$Z_i$ and independent with $Z_j$, and $P_1/P_4$ is an increasing
function of $Z_i$ and $Z_j$.
Since
\[
P_1=\frac{P_1}{P_1+P_2+P_3+P_4}=\frac{1}{1+{P_2}/{P_1}+
{P_3}/{P_1}+{P_4}/{P_1}},
\]
thus $P_1$ is an increasing function of $Z_i$ and $Z_j$.
\end{pf}

Lemma~\ref{lemma:2} suggests that a larger value of $Z$ indicates a
larger probability of being a risk node. The probability of being a
risk node is an increasing function of the $Z$ score, given the risk
status of other nodes are fixed.

\section{Simulation}
\label{sec:simu}
In this section we use simulated data to evaluate our proposed models
and algorithms and
demonstrate the efficacy of our proposed method. We simulate $Z$-scores
and hidden states from the HMRF model as given in
(\ref{model}) and~(\ref{ising}). The gene expression levels are
simulated from a multivariate Gaussian distribution. First, we compare
the proposed PNS algorithm with other existing high-dimensional graph
estimation algorithms. Second, we compare the power to detect the risk
genes using graphs estimated using a variety of graph estimation
algorithms. Our objective is to determine if we can achieve better risk
gene detection when we incorporate the network estimated by PNS into
the HMRF risk gene detection procedure. This comparison also sheds
light on the advantages of DAWN relative to DAWN$_\alpha$.

\subsection{Data generation}\label{sec4.1}
We adopt the B--A algorithm [\citet{BA:1999}] to simulate a scale-free
network $G=(V,\Omega)$, where $V$ represents the list of nodes and the
adjacency matrix of the network is denoted as $\Omega$. To obtain a
positive definite precision matrix supported on the simulated network
$\Omega$, the smallest eigenvalue $e$ of $v\Omega$ is first computed,
where $v$ is a chosen positive constant. We then set the precision
matrix to be $v\Omega+(|e|+u)\mathbf{I}_{d\times d}$, where $\mathbf
{I}_{d\times d}$ is the identity matrix, $d$ is the number of nodes,
and $u$ is another chosen positive number. Two constants $v$ and $u$
are set as 0.9 and 0.1 in our simulation. Finally, by inverting the
precision matrix, we obtain the covariance matrix $\Sigma$. Gene
expression levels, $X_1,\ldots, X_n$, are generated
independently from $N(0,\Sigma)$. The sample size $n$ is equal to 180
in our simulation.

To simulate $\mathbf{Z}$ from (\ref{model}), we first simulate the
hidden states $\mathbf{I}$ from the Ising model (\ref{ising}). Initial
values of $\mathbf{I}$ are randomly assigned to each node in the
simulated graph and we let half of the nodes have initial values
$I_i=1$. Then, we apply the standard Metropolis--Hastings algorithm to
update ${\mathbf I}$ with 200 iterations. 
The parameters in the Ising model (\ref{ising}) are set as $b=-7$ and
$c=3$ in our simulation.

\begin{figure}

\includegraphics{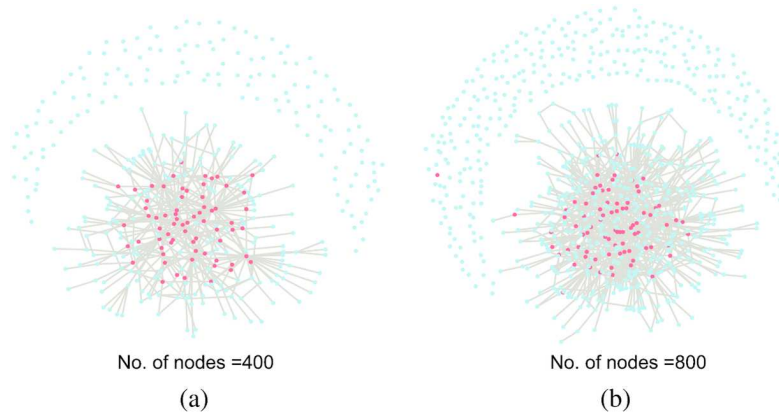}

\caption{Simulated scale-free network. \textup{(a)} number of nodes equals 400,
\textup{(b)} number of nodes equals 800.}
\label{fig:network}
\end{figure}

Figure~\ref{fig:network} shows the generated scale-free network with
the hidden states simulated from the Ising model. The numbers of nodes
$d$ are set at 400 and 800, respectively. In Figure~\ref{fig:network}(a)
there are in total 68 nodes with $I_i=1$ and in Figure~\ref{fig:network}(b) there are in total 82 nodes with $I_i=1$. After the
network and the hidden states embedded in the network are obtained, we
simulate $z$-score $Z_i$ based on model (\ref{model}) with $\mu=1.5$,
$\sigma_0=1$ and $\sigma_1=1$.

\subsection{Estimation and evaluation}\label{sec:simulation_evaluation}
Using the simulated data $X_1,\ldots, X_n$, we first estimate the graph
with the PNS algorithm. The $p$-value threshold $t$ is chosen to be 0.1
and the correlation threshold $\tau$ is set at 0.1. Define the
important edges as those edges connecting risk nodes. To evaluate the
performance of the PNS algorithm in retrieving important edges, we
compare the following three graph estimation algorithms:
\begin{itemize}
\item PNS: The proposed PNS algorithm.
\item Glasso: Graphical lasso algorithm.
\item Correlation: Compute the pairwise correlation matrix $M$ from
$X_1,\ldots, X_n$, then estimate graph $\Omega_{ij}=I\{|M_{ij}|>\tau\}$.
\end{itemize}

\begin{figure}

\includegraphics{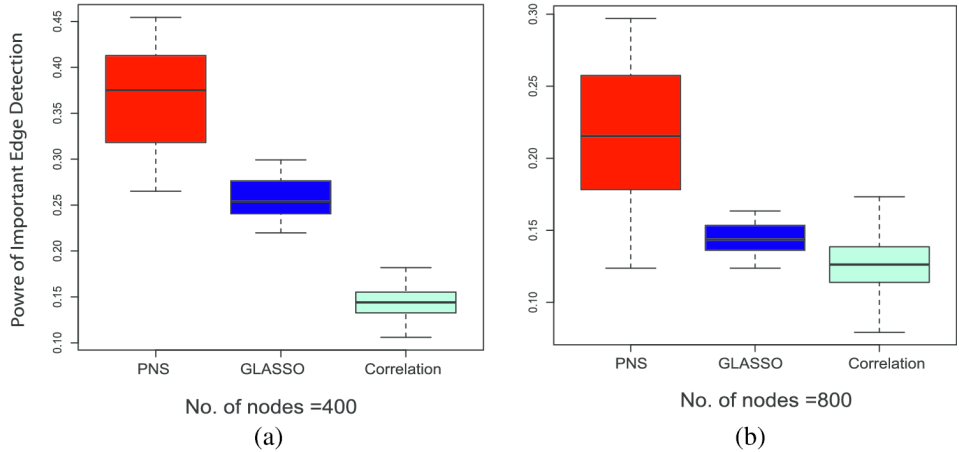}

\caption{Power of important edge detection. The FDR of the three
approaches is set at 0.5.}
\label{fig:edge}
\end{figure}

To compare the performance of graph estimation, the FDR is defined as
the proportion of false edges among all the called edges. Power is
defined as the proportion of true, important edges that are called
among all the important edges in the true graph. Figure~\ref{fig:edge}
shows that under the same FDR, PNS retrieves many more important edges
than the Glasso and Correlation algorithms. Calling more true edges
between risk nodes will improve performance, but calling more false
edges will reduce the power of the HMRF algorithm. From the comparison
in Figure~\ref{fig:edge}, we see that when calling the same number of
false edges, the PNS algorithm calls more true important edges, which
suggests that the HMRF model can achieve better power when using the
network estimated by the PNS algorithm. We will examine this conjecture
by comparing the power of the HRMF model using networks estimated with
different algorithms. The tuning parameters for each model are chosen
to yield a preset FDR. It is worth noting that here the PNS algorithm
does not use the scale-free criterion to choose the sparsity parameter
$\lambda$. Thus, the good performance of PNS does not really depend on
the scale-free assumption.

To evaluate the power of network assisted risk gene detection, we apply
the HMRF model using an estimated graph $\hat{\Omega}$ and the
simulated $z$-score $\mathbf{Z}$. We compare the following four approaches:
\begin{itemize}
\item HMRF\_PNS: Apply the HMRF algorithm by incorporating the graph
estimated by PNS.

\item HMRF\_Glasso: Apply the HMRF algorithm by incorporating the graph
estimated by Glasso. The tuning parameter of Glasso is chosen to make
the estimated graph having the same number of edges with $\hat{\Omega
}$, the network estimated by PNS.

\item HMRF\_oracle: Apply the HMRF algorithm by incorporating the true
graph $\Omega$.

\item Naive: Classify the nodes only based on the observed $z$-score
$\mathbf{Z}$.
\end{itemize}

\begin{figure}

\includegraphics{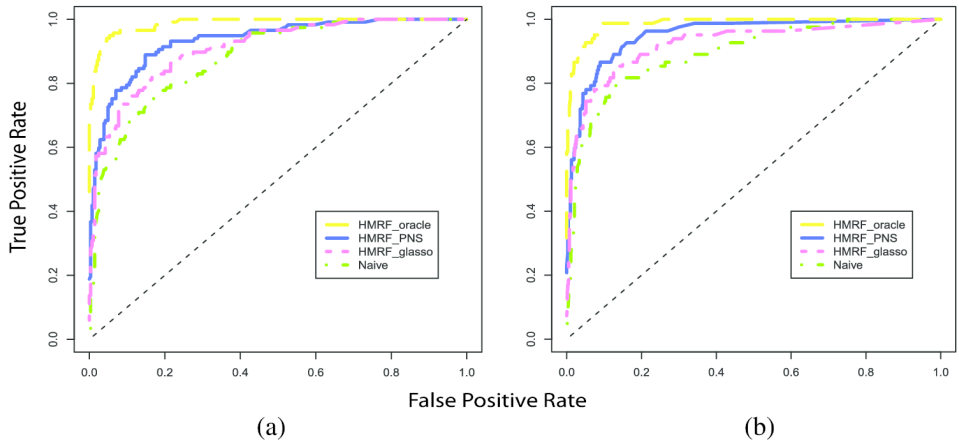}

\caption{ROC curve. \textup{(a)} number of nodes equals 400, \textup{(b)} number of nodes
equals 800.}
\label{fig:roc}
\end{figure}

Figure~\ref{fig:roc} shows the receiver operating characteristic (ROC)
curve of the four approaches applied to a single data set. We see that
by applying the HMRF model to incorporate the structural information
via the PNS algorithm, the accuracy rate of classification can be
largely improved. To evaluate the robustness of the proposed algorithm,
we repeat the simulation 20 times and compare the true positive rates
(TPR) obtained from each approach under the same false positive rate.
From Table~\ref{tab:res1}, we reach the same conclusion that HRMF\_PNS
performs much better than DAWN$\_{\alpha}$, HMRF\_Glasso and the Naive method.

%
\begin{table}[b]
\tablewidth=250pt
\caption{True positive rate comparison. The false positive rates are
controlled at 0.1}\label{tab:res1}
\begin{tabular*}{250pt}{@{\extracolsep{\fill}}lcc@{}}
\hline
&$\bolds{d=400}$ & $\bolds{d=800}$ \\
\hline
DAWN & 0.733 (0.02) & 0.732 (0.02) \\
DAWN$_\alpha$ & 0.663 (0.02) & 0.612 (0.03) \\
HMRF\_Glasso & 0.670 (0.03) & 0.651 (0.01) \\
HMRF\_Oracle &0.934 (0.02)& 0.917 (0.01)\\
Naive& 0.585 (0.02)& 0.567 (0.01)\\
\hline
\end{tabular*}
\end{table}

This simulation experiment also yields insights into advantages of
DAWN over DAWN$_\alpha$. A key difference between the algorithms is
that DAWN$_\alpha$ utilizes an estimated correlation network, while
DAWN relies on the PNS partial correlation network. Comparing the two
approaches in Figures~\ref{fig:edge}, \ref{fig:roc} and Table~\ref{tab:res1} reveals notable differences. It
appears that the correlation network fails to capture a sizable portion
of the correct edges of the graph. Consequently, the HMRF has a greater
challenge discovering the clustered signal. Overall, the simulations
suggest that DAWN performs much better because it uses PNS to fit the graph.

Next, we examine the robustness of our proposed DAWN under different
tuning parameters. To generate Table~\ref{tab:res1}, we chose $t=0.1$
and $\tau=0.1$. Now, we vary the tuning parameters $t$ and $\tau$ and
reevaluate the performance of DAWN. For $t$ we use five different
values $0.06,0.08,0.1,0.12,0.14$, and for $\tau$ we use three different values
$0.05,0.1,0.15$. The comparison is made using the same 20 simulated
data sets that were used to generate Table~\ref{tab:res1} (node${}={}$800).

%
\begin{table}
\caption{True positive rate comparison of DAWN under different parameters}\label{tab:res2}
\begin{tabular*}{\textwidth}{@{\extracolsep{\fill}}lccccc@{}}
\hline
&$\bolds{t=0.06}$ &$\bolds{t=0.08}$ &$\bolds{t=0.1}$ &$\bolds{t=0.12}$ &$\bolds{t=0.14}$ \\
\hline
$\tau=0.05$& 0.665 (0.01) & 0.709 (0.01) & 0.732 (0.02) & 0.717 (0.01)
& 0.699 (0.01) \\
$\tau=0.1$ & 0.666 (0.01) & 0.709 (0.01) & 0.732 (0.02) & 0.717 (0.01)
& 0.698 (0.01) \\
$\tau=0.15$ & 0.667 (0.01) & 0.708 (0.01) & 0.731 (0.02) & 0.717 (0.01)
& 0.698 (0.01) \\
\hline
\end{tabular*}
\end{table}

From Table~\ref{tab:res2} we see that the results of DAWN are not
sensitive to the choice of $\tau$. The tuning parameter $t$ does affect
the performance of our algorithm. If $t$ is too small, we will not have
enough seed genes for constructing the network and too many pairs of
key genes are missed.
But as long as $t$ is not too small, the performance of our algorithm
is robust. Hence, it is reasonable to choose a $t$ that is not too
small because in the screening stage we prefer overinclusion. Finally,
comparing Tables~\ref{tab:res1} and~\ref{tab:res2}, we see that
for every combination of parameters, DAWN outperforms DAWN$_\alpha$ and
HMRF\_glasso.

\section{Analysis of autism data}
\label{sec:data}
Building on the ideas described in Section~\ref{sec:Background}, Background and Data, we
search for genes association with risk for autism.
The gene expression data we use to estimate the network was produced
and normalized by \citet{Kang:2011}. \citet{Willsey:2013} identified the
spatial/temporal choices crucial to neuron development and highly
associated with autism. Networks were estimated from the FC during
post-conception weeks 10--19 (early fetal) and 13--24 (mid fetal). Thus,
we apply PNS to estimate the gene network using brains in early FC and
mid FC, respectively. For a given time period, all corresponding tissue
samples were utilized. In the early FC period there are 140
observations and in the mid FC period there are 107 observations. To
represent genetic association TADA $p$-values, $p_i$ are obtained from
\citet{DeRubeis:2014} for each of the genes.

\begin{figure}

\includegraphics{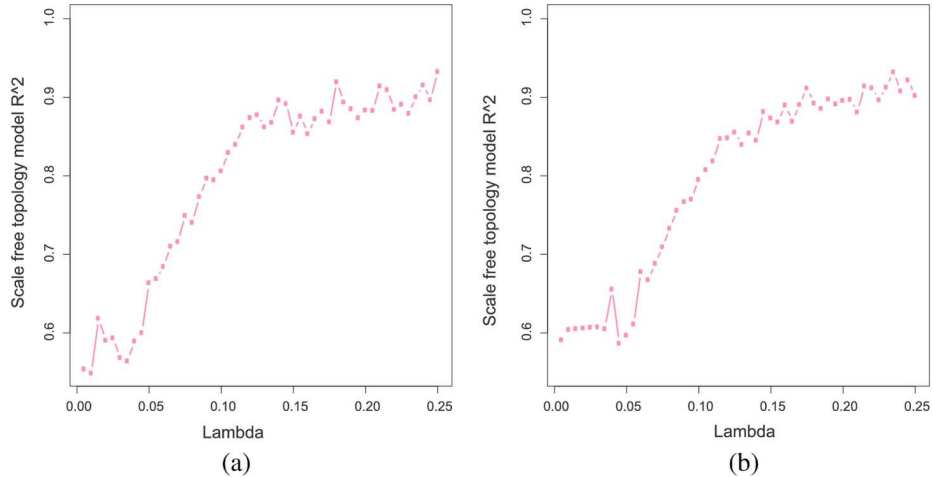}

\caption{Scale-free topology criteria. \textup{(a)} early FC \textup{(b)} mid FC.}
\label{fig:scale}
\end{figure}

\begin{figure}[t]

\includegraphics{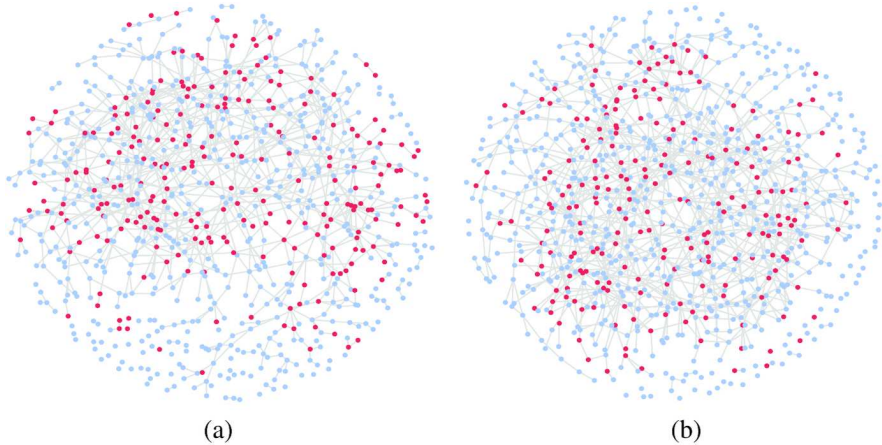}

\caption{Result of HMRF\_PNS algorithm for autism data. \textup{(a)} early FC
\textup{(b)} mid FC.}
\label{fig:realdata}
\end{figure}

The PNS algorithm is applied to early FC and mid FC separately. The
$p$-value threshold $t$ is chosen to be 0.1 and the correlation threshold
$\tau$ is set as 0.1. After the screening step, in early FC there are
6670 genes of which 834 genes have $p$-values less than 0.1, and in mid
FC there are 7111 genes of which 897 genes have $p$-value less than 0.1.
We define these genes with $p$-value less than 0.1 as key genes. To
choose the tuning parameter $\lambda$, we apply the scale-free criteria
and plot the square of correlation $R^2$ [equation (\ref{rsquare})]
versus $\lambda$ in Figure~\ref{fig:scale}. Based on the figure we
select, $\lambda=0.12$ because it yields a reasonably high $R^2$ value
in both periods. The full network of all analyzed genes in early FC
contains 10,065 edges of which 1005 edges are between key genes, and the
subnetwork of key genes is shown in Figure~\ref{fig:realdata}(a). The
full network of all analyzed genes in mid FC contains 11,713 edges of
which 1144 edges are between key genes, and the subnetwork of key genes
is shown in Figure~\ref{fig:realdata}(b).

After the networks are estimated, we assign $z$-scores to each node of
the network and then apply the HMRF model to the network. The initial
hidden states of genes are set as $I\{p_i<0.05\}$. We fix the hidden
states of 8 known autism genes as~1. Those 8 known autism genes are
\textit{ANK2\textup{,} CHD8\textup{,} CUL3\textup{,}
DYRK1A\textup{,} GRIN2B\textup{,} POGZ\textup{,} SCN2A} and \textit{TBR1}
based on \citet{Willsey:2013}. We then compute the Bayesian FDR value
[\citet{muller:2006}] of each gene based on the posterior probability
$q_i$ obtained from the HMRF algorithm. Under the FDR level of 0.1, we
obtain 246 significant genes in early FC of which 114 have at least one
identified dnLoF mutation. In mid FC we obtain 218 significant genes of
which 115 have at least one dnLoF mutation. We combine the significant
genes from those two periods and obtain in total 333 genes as our final
risk gene list (Supplemental Table~1 [\citet{supp}]). Among them, 146 genes have at
least one dnLoF mutation. Comparing to the number of genes discovered
by TADA [\citet{DeRubeis:2014}] where structural information of the
genes was not incorporated, the power of risk gene detection has been
substantially improved. The genes in the risk gene list are red in
Figure~\ref{fig:realdata}. From the figure it is clear that those genes
in the risk gene list are highly clustered in the network.

In our risk gene list, in addition to the 8 known ASD genes, there are
10 additional genes that have been identified as ASD genes [\citet
{Betancur:2011}] (three syndromic: \textit{L1CAM\textup{,} PTEN\textup{,} STXBP1}; two with
strong support from copy number and sequence studies: \textit{MBD5\textup{,}
SHANK2}; and five with equivocal evidence: \textit{FOXG1\textup{,}
FOXP1\textup{,} NRXN1\textup{,}
SCN1A\textup{,} SYNGAP1}). Fisher's exact test shows significant enrichment for
nominal ASD genes in our risk gene list ($p$-value${} = 2.9\times
10^{-6}$).

Next we compare the performance of DAWN$_\alpha$ and DAWN on the
autism data. Ranking the DAWN$_\alpha$ genes by FDR $q$-value, we retain
the top 333 genes for comparison. Autism risk genes are believed to be
enriched for histone-modifier and chromatin-remodeling pathways [\citet
{DeRubeis:2014}]. Comparing the DAWN$_\alpha$ and DAWN gene list with
the 152 genes with histone-related domains, we find 9 of these
designated genes are on the DAWN$_\alpha$ list (Fisher's exact test
$p$-value${} = 4.7\times10^{-2}$) and 11 are on the DAWN list ($p$-value${} =
5.5\times10^{-3}$). Thus, DAWN lends stronger support for the
histone-hypothesis and, assuming the theory is correct, it suggests
that DAWN provides greater biological insights, but this does not prove
that DAWN is better at identifying autism risk genes. Using new data
from \citet{Iossifov:2014}, we can conduct a powerful validation
experiment. Summarizing the findings from the 1643 additional trios
sequenced in this study, we find 251 genes that have one or more
additional dnLoF mutations. Based on previous studies of the
distribution of dnLoF mutations, we know that a substantial fraction of
these genes are likely autism genes [\citet{Sanders:2012}]. We find 18
and 24 of these genes are in the DAWN$_\alpha$ and DAWN lists,
respectively. If we randomly select 333 genes from the full genome, on
average, we expect to sample only 4--5 of the 251 genes. Thus, both
lists are highly enriched with these probable autism genes (Fisher's
exact test $p$-value${} = 2.4\times10^{-6}$ and $3.4\times10^{-10}$,
resp.). From this comparison we conclude that while both models
are successful at identifying autism risk genes, DAWN is more powerful.

We further investigate the robustness of our model to the lasso tuning
parameter, $\lambda$, by comparing the risk gene prediction set using
two additional choices bracketing our original selection. We identified
324, 333 and 243 risk genes with FDR $<0.1$ using $\lambda=0.10$, 0.12
and 0.15, respectively. Not surprisingly, the gene lists varied
somewhat due to the strong dependence of the model on the estimated
network; however, overlap between the first and second list was 281,
and overlap between the second and third list was 197. The median TADA
$p$-value for risk genes identified was approximately 0.01 for each
choice of $\lambda$, suggesting the models were selecting genes of
similar genetic information on average. But the model fitted with the
strictest smoothing penalty ($\lambda=0.15$) identified a smaller
number of genes, and yet it retained some genes with weaker TADA
signals (95th percentile TADA $p$-value 0.3 versus 0.1 for the other
smoothing values). This suggests that there might be greater harm in
over-smoothing than under-smoothing. \citet{DeRubeis:2014} identified
107 promising genes based on marginal genetics scores alone (TADA
scores with FDR $< 0.3$), hence, we also examined consistency of the
estimators over this smaller list of likely ASD risk genes. Of these
genes, 12 of them do not have gene expression data at this period of
brain development and cannot be included in our analysis, reducing our
comparison for potential overlap to 95 genes. For the 3 levels of
tuning parameters, DAWN identified 82, 82 and 75 genes from this list,
respectively. We conclude that although the total number of genes
varies, the genes with the strongest signals are almost all captured by
DAWN regardless of the tuning parameter chosen. Nevertheless, to obtain
a more robust list of risk genes, it might be advisable to use the
intersection of genes identified by a range of tuning parameters.

\begin{figure}

\includegraphics{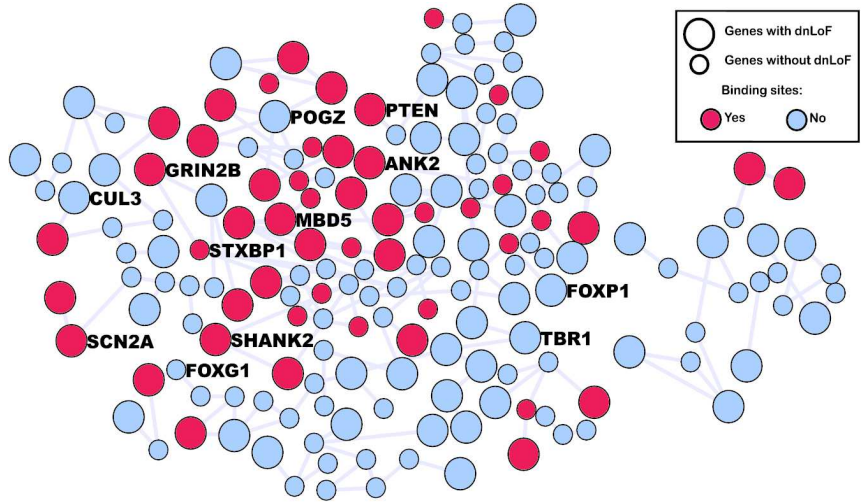}

\caption{Risk genes identified after incorporating FMRP targets.}
\label{fig:chd8network}
\end{figure}

The Ising model allows for the incorporation of numerous covariates
such as TF binding sites either individually or en masse. To
illustrate, we incorporate the additional information from targets of
\textit{FMRP} [\citet{darnell:2011}]. These target genes have been shown to
be associated with autism [\citet{Iossifov:2012}], hence, it is
reasonable to conjecture that this covariate might improve the power of
autism risk gene detection. Indeed, the additional term is significant
in the Ising model ($p < 0.005$ obtained from Algorithm \ref{alg:sim}
and $B$ is set as 200). Applying model (\ref{ising}) to the early FC
period, we discovered 242 genes of which 118 have at least one dnLoF
mutation. Four of the genes with one dnLoF mutation are newly
discovered after we incorporate the TF information. Those four genes
are \textit{TRIP12\textup{,} RIMBP2\textup{,} ZNF462} and
\textit{ZNF238}. Figure~\ref{fig:chd8network} shows the connectivity of risk genes after
incorporating the TF information.

\section{Conclusion and discussion}\label{sec6}
In this paper we propose a novel framework for network assisted genetic
association analysis. The contributions of this framework are as
follows: first, the PNS algorithm utilizes the node specific
information so that the accuracy of network estimation can be greatly
improved; second, this framework provides a systematic approach for
combining the estimated gene network and individual genetic scores;
third, the framework can efficiently incorporate additional structural
information concerning the dependence between genes, such as the
targets of key TFs.\looseness=1

A key insight arises in our comparison of the HMRF model using a
variety of network estimation procedures. The Glasso approach tries to
reconstruct the whole network, while the PNS approach focuses on
estimating only the portions of the network that capture the dependence
between disease-associated genes. It might seem counterintuitive that
the HMRF model can achieve better power when the network is estimated
by the PNS algorithm rather than by other existing high-dimensional
network estimation approaches such as Glasso. Why would we gain better
power when giving up much of the structural information? Results using
the oracle show that HMRF works best when provided with the complete
and accurate network (Table~\ref{tab:res1}). The challenge in the
high-dimensional setting is that it\vadjust{\goodbreak} is infeasible to estimate the
entire network successfully. Hence, the PNS strategy of focusing effort
on the key portions of the network is superior. With this approach more
key edges are estimated correctly relative to the number of false edges
incorporated into the network.

While we build on ideas developed in the DAWN$_{\alpha}$ model [\citet
{Liu:2014}], the approach presented here extends and improves
DAWN$_{\alpha}$ in several critical directions. In the original
DAWN$_\alpha$ model, the gene network was estimated from the adjacency
matrix obtained by thresholding the correlation matrix. To obtain a
sparse network, DAWN$_\alpha$ grouped tightly correlated genes together
into multi-gene supernodes. DAWN uses PNS to obtain a sparse network
directly without the need for supernodes. This focused network permits
a number of improvements in DAWN. Because each node in the network
produced by the PNS algorithm corresponds to a single gene, it is
possible to directly apply the Bayesian FDR approach to determine risk
genes. In contrast, the DAWN$_\alpha$ required a second screening of
genes based on $p$-values to determine risk genes after the HMRF step.
Finally, DAWN is more flexible and allows for the incorporation of
other covariates into the model.

The proposed framework is feasible under different scenarios and has a
wide application in various problems. In this paper, we extended the
Ising model so that the proposed network assisted analysis framework
can be applied to incorporate both the gene co-expression network and
the gene regulation network. This framework can also be naturally
extended to incorporate the PPI network together with the gene
co-expression network by simply adding another parameter in the Ising
model. These three different types of networks can even be integrated
simultaneously to maximize the power of risk gene detection. Moreover,
the proposed risk gene discovery framework can be applied not only to
ASD but also to many other complex disorders.

\section*{Acknowledgments}

We thank the Autism Sequencing Consortium for compiling the data, and
Bernie Devlin, Lambertus Klei and Xin He for helpful comments.


\begin{supplement}[id=suppA]
\stitle{Supplemental Table~1: Statistics for all genes analyzed in early and mid FC periods}
\slink[doi]{10.1214/15-AOAS844SUPP} 
\sdatatype{.zip}
\sfilename{aoas844\_supp.zip}
\sdescription{Column min\_FDR is the minimum value of FDR of both periods.
For the risk\_early and risk\_mid columns, a gene was labeled 1 if it
was identified.
FDR\_early and FDR\_mid column report the FDR value of each gene in
early and mid FC periods.
The dn.LoF column is the number of identified dnLoF mutations in each gene.
The $p$-value column is the TADA $p$-value.}
\end{supplement}





\printaddresses
\end{document}